# Fluids in motion: Contemporary artScience- inspiration and realization

***Norman J. Zabusky***

Department of Physics of Complex Systems , Weizmann Institute of Science

Rehovot, Israel

 

"The creative act is not performed by the artist alone; the spectator brings the work in contact with the external world by deciphering and interpreting its inner qualifications and thus adds his contribution to the creative act". (Marcel Duchamp)

"The process of understanding in nature, together with the joy that man feels in understanding , i.e., in becoming acquainted with new knowledge, seems therefore to rest upon a correspondence, a coming into congruence of preexistent internal images of the human psyche with external objects and their behavior. ... the place of clear concepts is taken by images of strongly emotional content, which are not thought but are seen pictorially, as it were, before the minds eye…" ( W. Pauli)

**ABSTRACT** : I examine contemporary work in fluids in motion and demonstrate strong connections between visual art and science resulting from innovative technology. In one burgeoning domain, falling liquid drops impacting liquid pools it is valuable to compare how artists and scientists describe their goals and their use of *high speed photography* to capture and measure events. I also examine the use of devices to create still-pictures and animations : computer-simulation and visualization and installations at focal locations. In particular, Ned Kahn whose public domain installations show various fluids in motion - water, sand, fog, fire, etc which exhibit turbulent boundary layers, vortex rings, whirlpools, waterfalls, etc. Finally, I examine aspects of the role of digital technology and its utilization by artists, educators , museums and galleries for innovative and interactive displays.

## 1. INTRODUCTION

As art historian Martin Kemp noted [1] , "To generalize about the relationship between art and science is not so much hazardous as impossible. Neither science nor art are homogeneous categories. ...we serve any enquiry into art and science badly if our criterion is superficially the influence of science on art, or the influence of art on science. Deeper realms of enquiry concern complex dialogues centered on issues of cognition, perception, intuition, mental and physical structures, the communicative and social action of images, and the role of what we call the aesthetic ..."

Few *practicing* scientists have written on the subject. Distinguished physicists W. Heisenberg [2] and S. Chandrasekhar [3] have contributed enormously to quantum mechanics and astrophysics, respectively. They have written and focused on the ideal of *beauty and aesthetics* in science and art. I will try to make this art-science relationship possible to *generalize* by focusing mainly on fluid motion and innovative technology, a subject of my research for many decades.

Historically, Leonardo da Vinci (1452 –1519) the Renaissance polymath ( i.e., painter, sculptor,

draftsman, scientist, engineer, inventor, anatomist, writer and more) was the first to draw and paint images across STEM [4] disciplines. His deep appreciation of vortex-and-turbulence fluid dynamics in diverse fluid environments is uncanny and he may be considered the "father" of flow visualization [5].

Leonardo's painting, "Natural Disaster " in the "Deluge" series is very realistic and we observe at least two sizes of the "curls" - a large scale circular curve in the lower half on which we see upward curling small scale curls, all evidence of a jet or vortex ring striking the ground from above and becoming unstable and turbulent. Is this something he imagined or was he inspired by a real event ? Stanley Gedzelman [6], a meteorologist, argued that Leonardo was attempting to show what we now call a "downburst". They are transient vertical jets from a rain cloud that are close to the earth's surface. In papers written in 1981 and 1985, Ted Fujita [7] showed that the crash of a landing flight at New York's Kennedy Airport in 1975 was due to a sudden downburst. So meteorologists have likely interpreted the source of the inspiration for this painting.

*Why fluid motion?* Much of the mass of the cosmos is a fluid of matter and dark matter, some at extreme conditions of temperature and density- a radiating plasma. Much of planetary environmental phenomena is the result of dynamic interactions of the liquid oceans and gases and vapors in the atmosphere. These interactions occasionally generate violent storms, waves, floods, fires, etc . Finally, living entities like insects, plants, animals and humans contain "complex " fluids whose motion on the scale of microns sustains life.

Many contemporary artistic realizations by artists and *stemists* , elicit interest and excitement because they control and project evolving coherent and chaotic patterns in varying spatiotemporal domains. Evolving technology has resulted in new and improved devices , including the optical laser, high-speed (HS) camera, space telescope and digital computer. These are providing massive amounts of data that are yielding remarkable insights for scientific innovation. Recently, "inexpensive" HS digital cameras have become available and both artists and stemists have been quick to utilize these. They can create still images and videos of *previously unseen phenomena*.

## 2. Where I am coming from

As an applied mathematician and natural scientist, I have been blessed with time to do fundamental and applied work in fluid-and-wave dynamics [8] . At Bell Labs in New Jersey and academia ('62-'06), I became a computational fluid scientist, and continued working with wizardly [9] younger colleagues. Few art and science essayists and interviewees bring this background and immersion to their communications.

For many years, beginning in Pittsburgh, I worked with a young artist, Hilary Langhorst Shames. Together we talked and collaborated on numerous posters, pictures and an abstract animation with music, *Cosmic Vortex Projectiles*[i] [10]. After several interactions she designed and made Fig. 1, my scientific *auto*montage. In 2005, I organized and contributed to the 4th *ScArt* international conference in New Brunswick, N.J. [11] . Several papers given at the meeting as well as talks I have given are linked and worth a visit. Soon, the remainder will be posted.

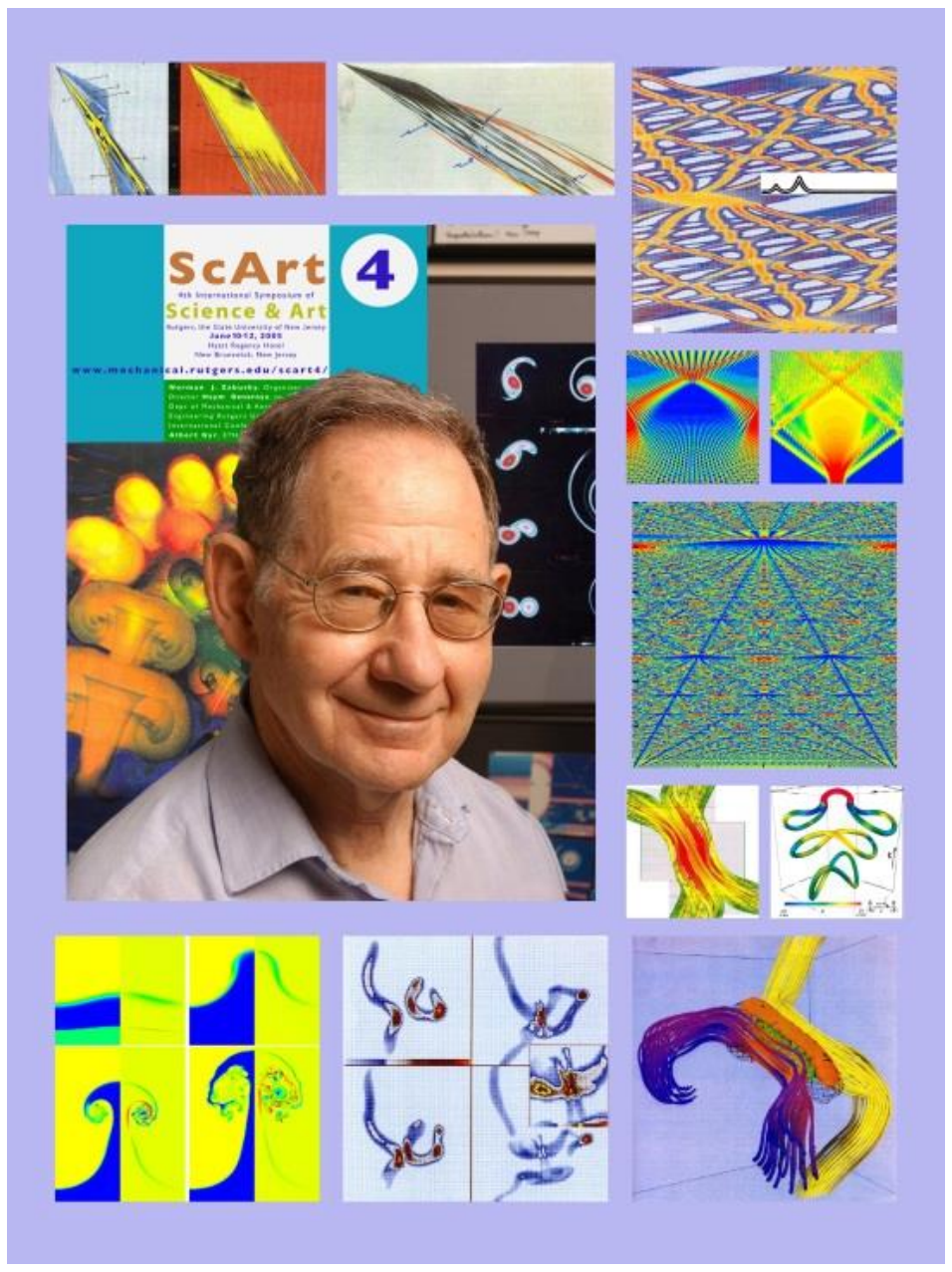

Fig. 1. N. J. Zabusky, Professional Photomontage by H. L. Shames including photo by Nick Romanenko/Rutgers University

Figure 1. by H. L. Shames contains a photo of me surrounded by panels of computer generated images and “projections” to lower dimensions [12] . Projection diagrams and other varieties are sources of intuition for innovation via the *visiometrics* mode of working [13]. All have been published in refereed physics and fluids journals.

## 3. **Photography**

To show how *art-science* realization depends on technology, let’s be specific and focus on two items introduced above : 1). high-speed and microscopic digital photography for complex fluids and 2). installation, performance and gallery/museum art emphasizing fluid motion.

After photography was invented the curious were inspired to look for applications. In 1872, Edwaerd Muybridge settled a hotly debated question with a single photographic negative showing a trotting horse *airborne* during the trot. By 1878, spurred on by Leland Stanford, Muybridge photographed a horse at a trot using multiple cameras.

He influenced: inventor-photographer Étienne-Jules Marey [14] who recorded the first series of live action photos with a single camera by his method of *chronophotography*. Here, on one negative, multiple images separated by small time intervals showed birds descending and landing, pole-vaulters in action, etc. Marcel Duchamp [15], who painted "Nude Descending a Staircase" acknowledged Marey’s influence. Thus, the artist has been influenced by the stemist. I knew of Duchamps and this picture, e.g. at the Philadelphia art museum before I found Marey on a visit to Musee-d’Orsay in Paris. I was *thrilled* to find this connection.

In the 1930’s Harold E. Edgerton engineer, artist, professor and entrepreneur improved and advocated stroboscopic and high speed photography [16] . He used an open shutter in a dark room illuminated by light flashes of one microsecond duration . He produced numerous photos- and two related to fluid dynamics are shown in Fig. 2.

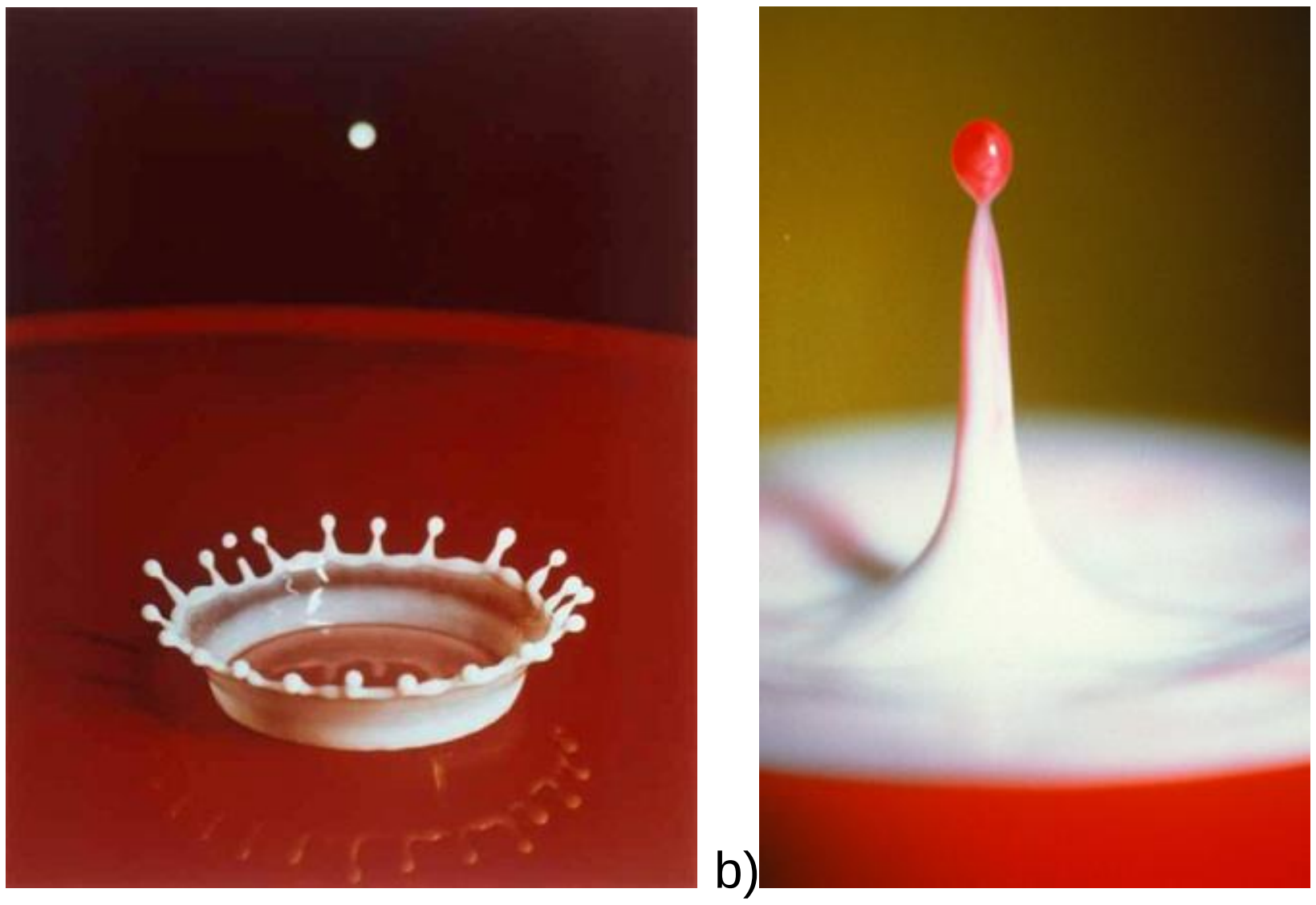

a) b)

Fig. 2. Some H. Edgerton *high-speed* images . a) Two milk drops have been released from a dropper. At this instant ,the upper small sphere is the second and is still falling; the first drop released has already impacted a thin milk layer resulting in an upward moving unstable “coronet” of nearly-equal spaced drops of milk. (b) A cranberry-red drop has been released and has impacted a pool of milk. At the instant shown the distorted and upward-moving red drop is followed by a white milk *stalk* .

The demands of experimental research and development in the 1980’s accelerated CMOS technology for sensors, high speed recording and computation. We now have cameras of up to 25 million frames per second (fps), with a typical high speed of one million fps. In the last few years relatively inexpensive and compact HS cameras, up to 5,000 fps have given photo-artists opportunities to enter new domains of space-time. They are using colored liquids or air in various controlled states to generate: breaking waves, sheets of water and distorted droplets; vapor/fog vortex rings; drops falling and colliding with solid surfaces, liquid pools or other drops in motion. The array of names includes *Marcel Christ, Jack Long, Vladimir Nefedov, Martin Waugh, Shinichi Maruyama* and surf photographer *Clark Little* [17].

Like scientists, the first four photograph carefully designed and instrumented space-time objects while *Maruyama* also hurls sheets of clear and colored water into space, as seen on stills and videos at his URL. *Little*, age 39, swims in terrifying seas and crouches on shorelines with his camera in water-housing equipment to capture rarely seen views of the sun from inside a 'tube' . His URL is a masterfully designed market of his pictures and objects.

Figure 3, includes one of many elegant HS photos by Vladimir Nefedov and one of many remarkable sunset photos of Clark Little.

3a) 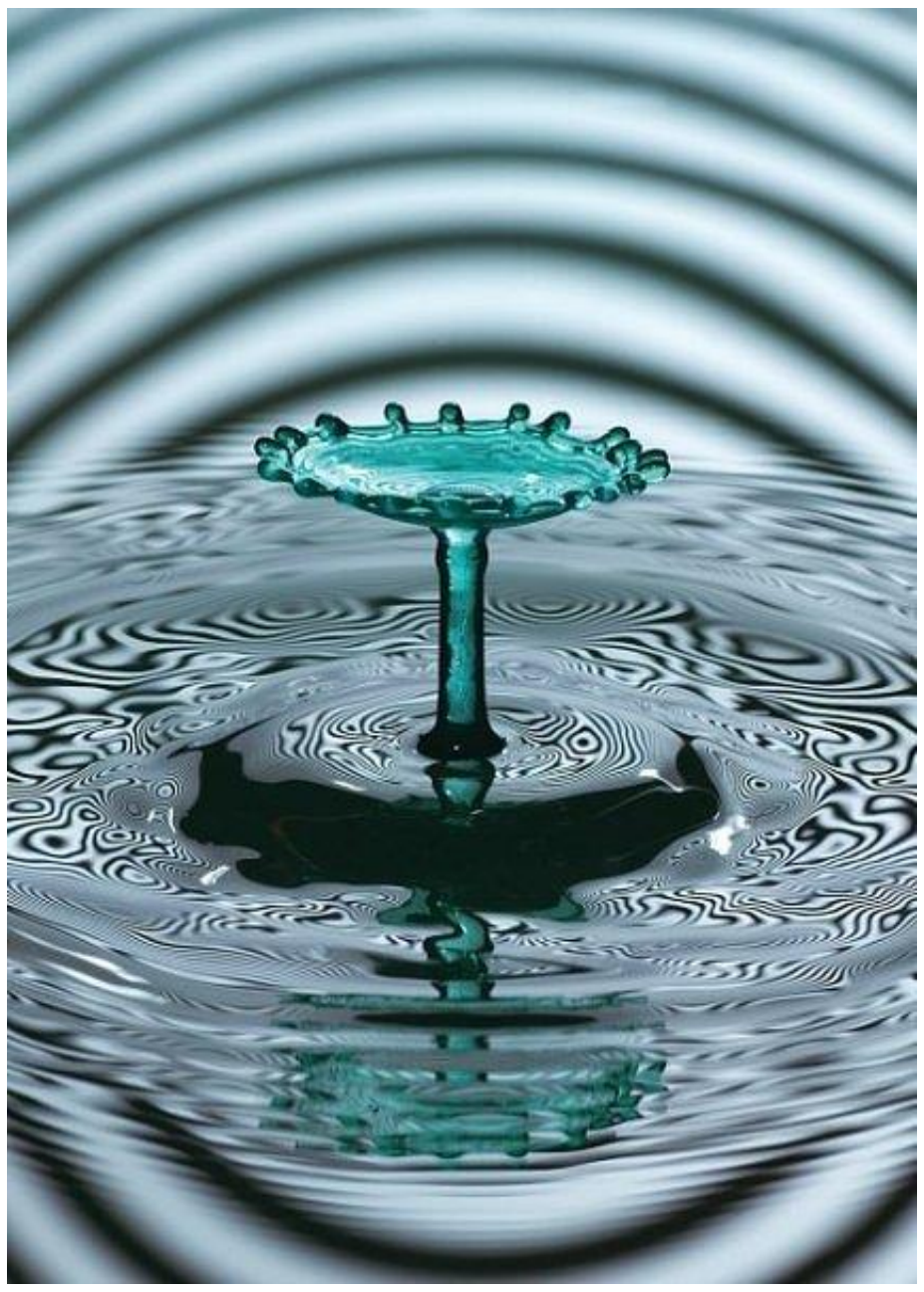 3b) 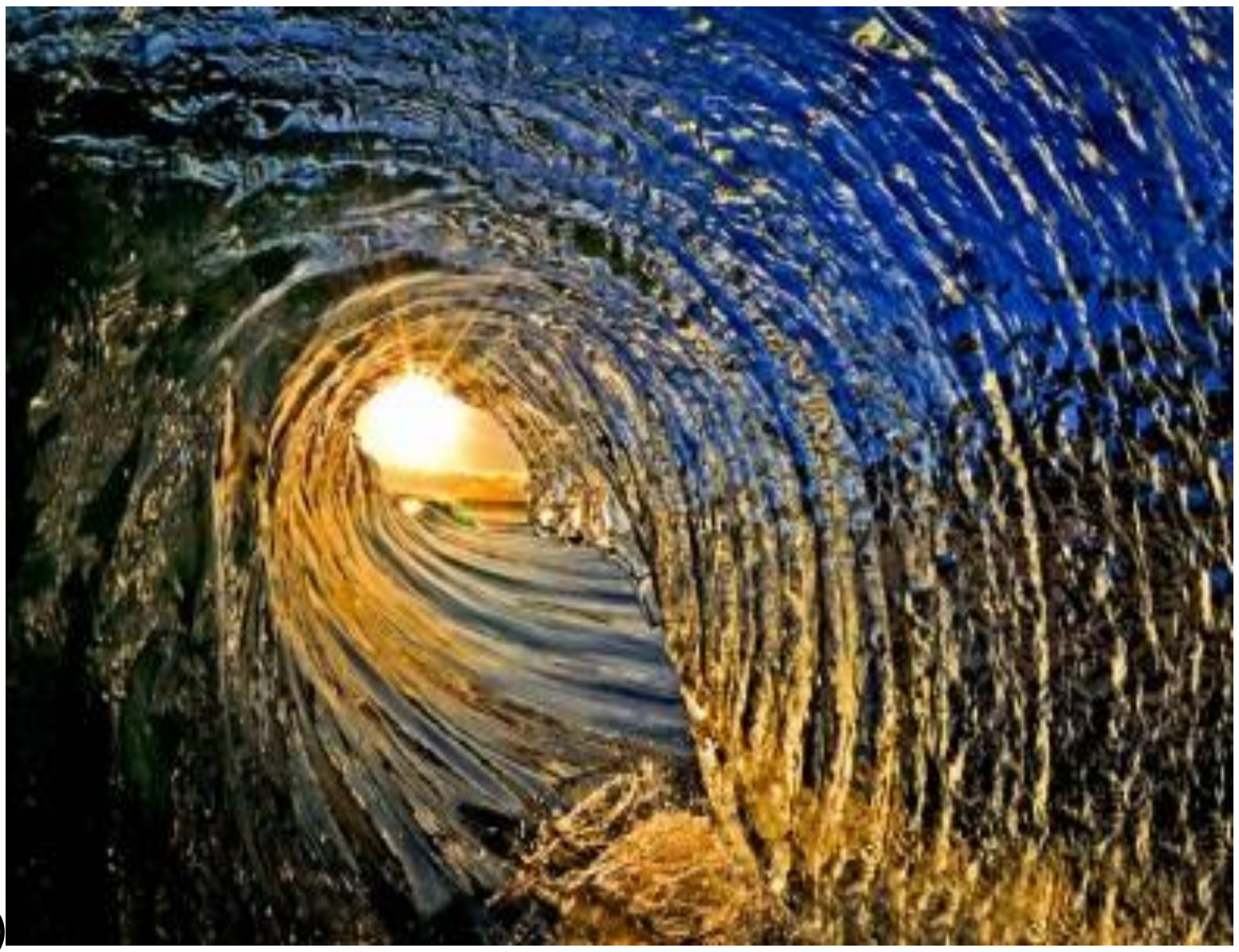

Fig. 3. a) Vladimir Nefedov- Drop impacts, carefully designed in space-time, produce a raised "coronet" with special coloring and effects. b) Clark Little -Sun setting through the "tunnel" of a breaking surf wave. Note the near-regular array of arcs on the falling water (right). They are evidence of an emerging fluid instability.

## 4. Scientific Experiments: Falling drops impacting liquid surfaces .4

Let's focus on a comparable contemporary stemist research domain: *liquid drops impacting horizontal solid and fluid surfaces.* This domain began in the late 19th century with English physicist Arthur Worthington's experiments using spark discharges [18]. There was little research activity until the early 2000's when better HS cameras became available . A contemporary review of the fluid dynamics appeared in 2006 [19] and since 2011 many publications, each with several authors have been published in high-impact journals.

Thorrodsen, Thoraval and six colleagues achieved a *juxtaposition* (experiment in black/white-not shown here- and axisymmetric simulation in four colors (R,Y,G,B) ) [20]. For a 4.6 mm diameter drop of water impacting a thin water layer, their simulations showed the appearance of bubble rings that are associated with dipolar vortex phenomenon *on* the radially moving interface. In the comprehensive follow-up paper [21] they explore this small region of space-time. They used a Shimadzu HS camera at *one million fps* and a resolution of 4.1 μm/pixel. Their *supplementary material* presents videos of the experiments as well as juxtaposed numerical simulations. Fig. 4 contains a terse sample provided by the authors.

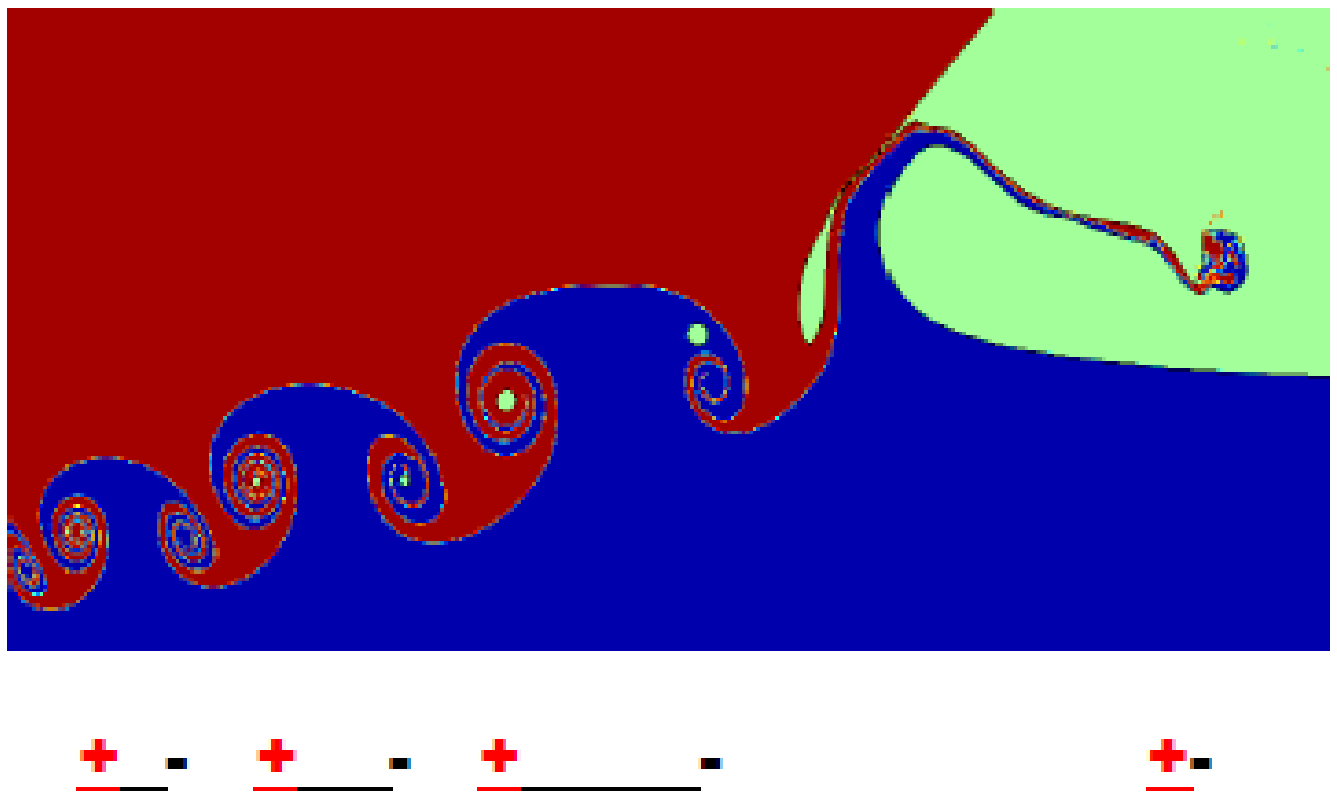

4a)

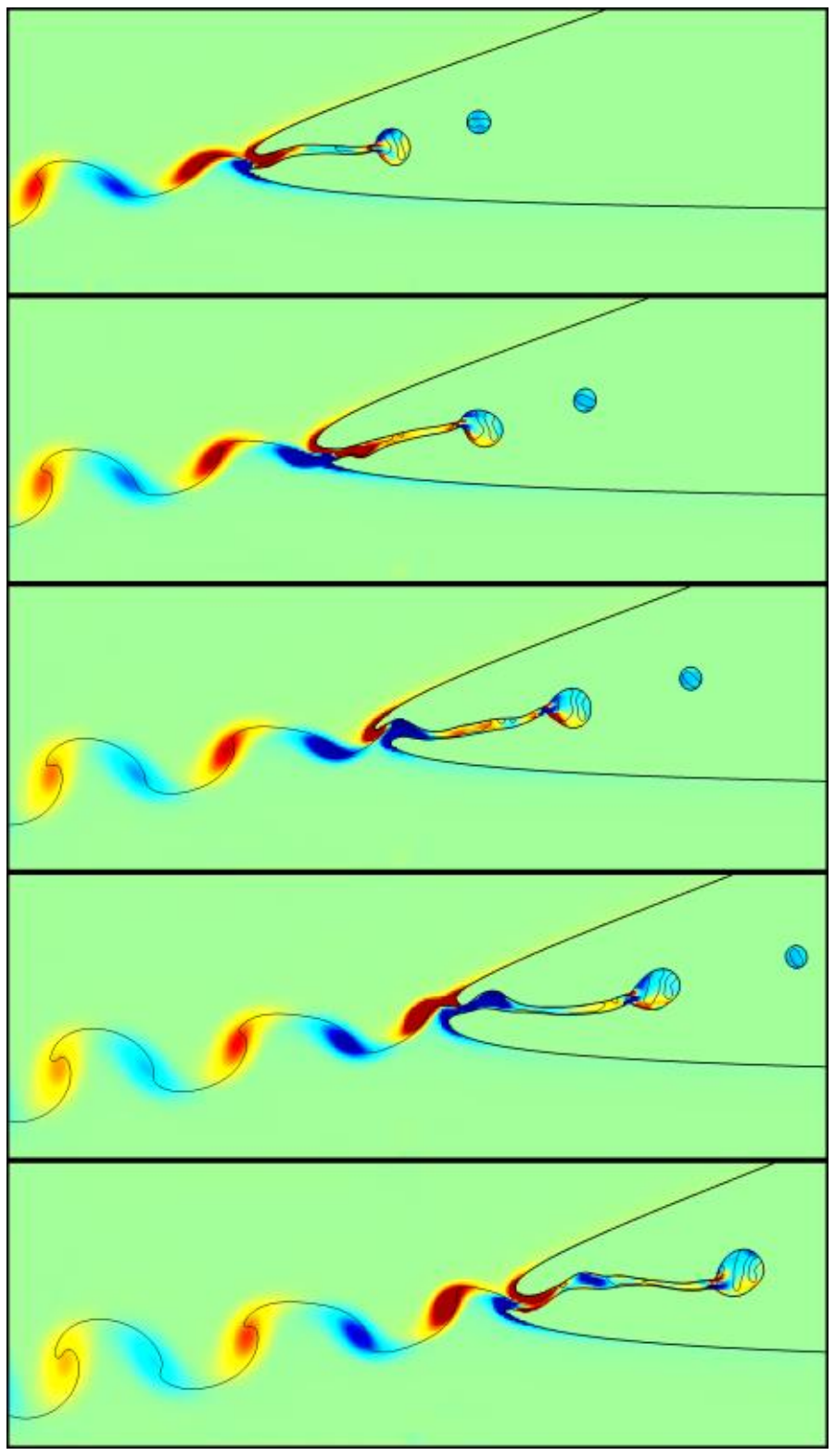

4b)

Fig.4. Emergence of entrapped "bubble rings" resulting from a water drop falling onto a horizontal water pool , from *axisymmetric simulations.* Note, D=4.6 mm and $t_{norm}$=(D/$U_{impact}$) =1.62 millisec so Re=14,500, We=617 and $K_{splash}$=74,400.

4a) A cropped image of a slice thru the axisymmetric simulation at t=<u>*107 µs*</u>. The colors indicate: *red*, the interior of the falling water drop, moving to the right and in contact with the *blue* previously horizontal undisturbed water pool; and *green* for the ambient *air.* A leading thin jet (containing mixed *red* and *blue* ) emerges ahead of the right-moving *red* domain. The trailing *red* and *blue* interface shows three of the many rolled-up "mushroom-shaped" domains on the interface as well as very small regions of entrained air ( *green)* . The distance between the third from left +__– above (within the red domain) is14.7 *µ*m or 0.32 % of the initial D= 4.6 mm.

4b). The *vorticity* at five times, separated by intervals of 1.1 µs; *red* is (+), *blue* is (--) and the black line is the evolving interface. The top panel shows a snapshot at t= <u>*19.9 µs*</u> after drop impact, where a red domain has been completely formed on the interface (fine black line) and to the right a *blue* domain is beginning to be formed*.* In the third panel the *blue* region has been completely formed and another *red* region is in process of being formed. Note the time of the top panel is before that shown in Fig. 4a .

Stemists seek to *scale* the behavior of physical phenomena. By *scale* I mean: how does some aspect of the evolving morphology change if we vary either Reynolds number (e.g.varying impact velocity, U, initial radius, $r_0$ or viscosity, $v$ in Re = $Ur_0/ v$ ) or Weber number (i.e. surface tension σ in We = $\rho U^2 r_0 / \sigma$ .) In the letter of 2012, the authors vary Re. When stemists do achieve a *new* level of understanding, they send their papers to high-impact journals. The artist wishes to capture unusual "artful" (or beautiful) and complex views, usually as colored still photographs, which he can exhibit in a gallery or display and sell on the internet, etc.

To summarize: when the artist and scientist examine or contemplate the same domain they seek different goals. The scientist tries to create the *simplest* spatial environment and discover new evolving morphologies, usually in the smallest spatial domain and for the shortest time interval. The artist seeks to minimally complexify the environment to obtain the picture or animation that captures patterns and colors that are artistically surprising and beautiful. This is evident in the simple colors from the juxtaposed numerical simulation that appear in Fig 4 when they are compared with, e.g., those of artists Vladimir Nefedov and Clark Little in Fig. 3.

## 5 . Performance and Installation of Kinetic-fluidic Art.

### A. Performance Fluidic Art

Paul Prudence' visual and sonic art accompanied by music fits well into the theme of this paper. The design and synchronization of the collective interactions are modernistic and I believe will provide another mode of artistic expression. His opus can be found at his URL [22] .

Here , I emphasize his "hydro-acoustic study." His Vimeo lecture is a combination of technical description followed by viewing-and-listening to the black and white animations. The juxtaposition of *sound* from water-immersed audio speakers and reflected *light* from lamps fixed in space and intensity shining on the surface waves is a curiously stimulating happening. When larger sound amplitudes will be used they should observe interesting *nonlinear* coherent and chaotic patterns.

### B. Installation Fluidic Art

In 2004, while preparing the program for the *ScArt4* meeting, I fortuitously "discovered" Ned Kahn , pioneering installation and kinetic-fluidic artist. He sent me a VCR of some of his work and after seeing it, I immediately invited him to participate. He accepted, but illness prevented him from coming. Ned has: worked in scientific settings (e.g.,the |San Francisco Exploratorium) ; decorated sides of numerous buildings with small aluminum flaps that react to the passing gusts of winds ; installed vortex ring generators in residential neighborhoods and commercial play and rest areas; museums, and etc . By occupying these cultural arenas simultaneously, his work and his ideas are interpreted within separate discourses – as educational, scientific demonstrations or as playful objects.

Asked whether his work is more science or art, he replies, "…they're definitely not scientific experiments, because they're often much more uncontrolled and complicated… On the other hand, they're not really artworks in the traditional sense… In the things that I make, even though I've created the physical structure, it's really not me that's doing the sculpting" [23]. In his 2003 MacArthur Award remarks "My artworks frequently incorporate flowing water, fog, sand and light to create complex and continually changing systems. …I am intrigued with the way patterns can emerge when things flow. These patterns are not static objects, they are patterns of behavior - recurring themes in nature".

Three of Kahn's large fluid dynamical installations are in different places at Singapore's Marina Bay Sands *urban forum and living center*, opened in 2011. The Center was created by architect Moshe Safdie who invited Ned to be the chief artist [24]. They are

1) "Wind Arbor" where a centrally located ambient-wind driven fine-structured vertical wall, exhibits randomly changing patterns;

2) "Rain Oculus", where a large swirling whirlpool ( about 70 feet in diameter) is created at street level ( Fig. 6a) and falls thru an oculus ( indented circular hole, Fig 6b) at 6,000 gallons/minute and descends two floors onto a pond as a rotating waterfall of bubbles and vortices (Fig.6c ); and

3) "Tipping Wall", where water at the top of the wall falls onto horizontal rows of mounted and pivoting rectangular plates and causes them to engage in a *dance of chaotic oscillations* and accompanying sounds.

These words hardly convey the unusual views and sounds to be seen in situ or in videos of these fluid dynamical environments.

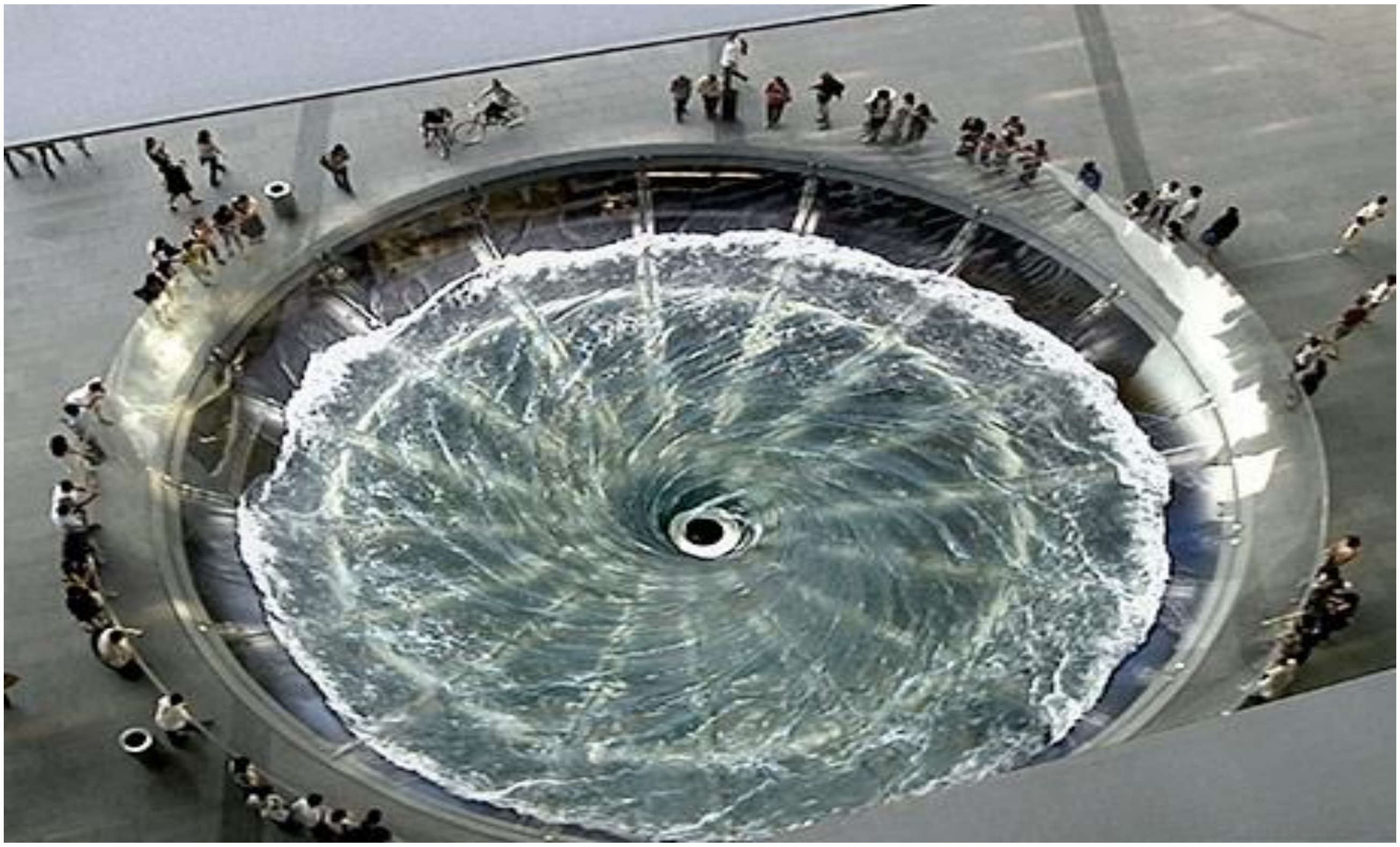

Fig 6a shows the street view of the *Rain Occulus, the* largest vortex-gyre in the world that may be generated at will. Fig 6b and Fig. 6c shows the below street level where the bubble-filled water “jet’ is falling through the occulus controllably and the panorama in Fig. 6c shows the jet spreading into the relaxation and shopping mall .

Marina Bay Sands has unique lotus shaped artScience building which contains a *museum* designed by the architect Safdie. Although it has a permanent exhibition, it mainly hosts touring exhibitions curated by other museums, all in 21 gallery spaces with a total area of 50,000 square feet (6,000 square meters. I look forward to my first visit!

c) 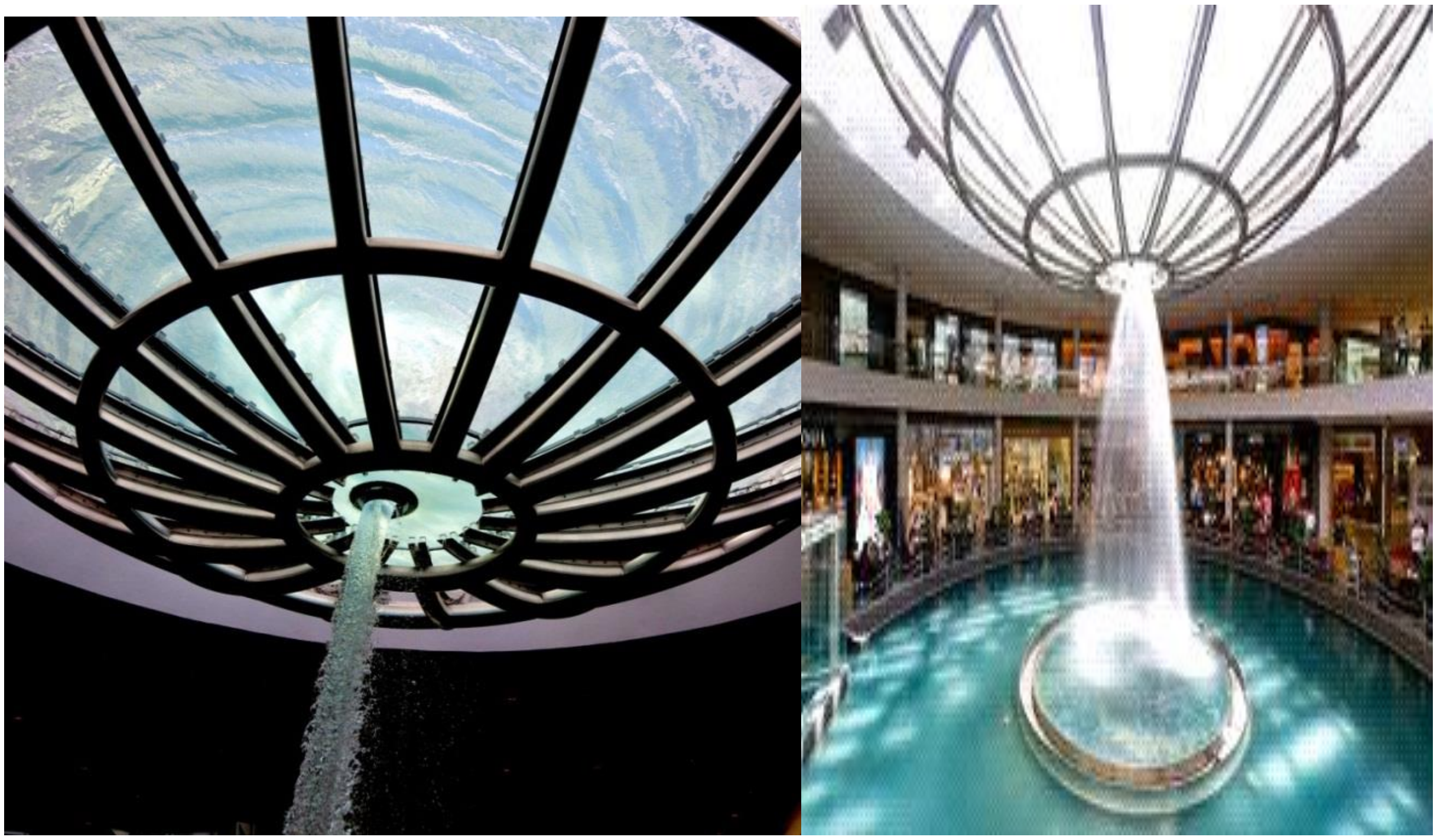

Fig. 6 "Rain Oculus" Three views with circulating water. (by Ned Kahn at Marina Bay Sands urban forum, 2011).

**6. Educating and Displaying Fluid Motion Art**

Jean Herzberg (teacher, fluid experimenter and technologist ) and Alex Sweetman (photographer) developed a course in flow visualization, “The physics and art of fluid flow” for undergraduates across disciplines [25]. It has run successfully since 2005 at the Boulder campus of the University of Colorodo. Perhaps it could be expanded into a degree program, say a Masters in Art, Design and Technology ?

Museums and galleries are now paying increasing attention to aspects of fluids-in-motion and technology. The Peabody Essex Museum’s interactive center (Salem , Massachusetts ) had a ten month exhibition, “*Ripple Effect, The Art of $H_2O$”* (June 2011 -April 2012). **W**ater as a source of artistic expression, featured 16 artists who represent water in its different states: solid, liquid, and gas. Staged in the Peabody’s Interactive Art and Nature Center, the show encourages viewers to interact with the art. Also, the exhibit *Surface Tension:The Future of Water* at the Science Gallery in Dublin ( September 2011) exhibited the effects of surface tension of water in various situations, e.g. the cone-shaped cavity arising on the surface of a water-filled rotating container. An internet search yields many more.

I expect that visitors to museums will enjoy diverse modes of *interactivity with art .* For example, Petros Vrellis has created an interactive visualization with synthesizer that animates Vincent Van Goghs “Starry Night ” [26]. A fluid *simulation* creates a flowing fabric mimicking Van Gogh’s impressionist portrait. A *touch interface* allows a viewer to deform the image, altering both the flow of the particles and the *synthesized sound*, and then watch it slowly return to its original state.

The sound itself is created using a MIDI interface to create a soft ambient tone out of the movement of the fluid.

## 7. Summary and Conclusion:

Using fluids in motion, I have attempted to show the motivations and results of artists and stemists with the new technology of hi-speed photography and with old and new venues. Mostly, they are younger people doing closely coupled science, design and art . Some have degrees in the arts , but rather more in science, computers, engineering, and environmental studies. Their works have been easily communicated on the internet through blogs and social networks and are being increasingly recognized in galleries, museums and marketing venues.

**Acknowledgements**. I appreciate the invitation by Professors Victor Steinberg and Shmuel Rubinstein to consult with students and faculty in their laboratories at the Weizmann Institute of Science. Steinberg , colleagues, post-docs and students have done pioneering experiments to establish transitions in the dynamics of vesicles in microfluidic shear flows. Rubinstein and colleagues at Harvard have done high speed experiments of liquid drop impacts on solid surfaces. I also benefitted from the diverse and informative comments made by reviewer Evelina Domnitch and from recommendations by Jeanne Herzberg to URL’s which emphasize fluid dynamics and art.

7. J. W. Wilson, and Roger M. Wakimoto "The Discovery of the Downburst - TT Fujita's Contribution". Bulletin of the American Meteorological Society, vol 82 (1). (2001).

8. I began as an Electrical Engineer and was introduced to the "best" in computing devices: MIT ('51-'53), the Whirlwind I digital computer; Caltech, Dept of Physics ('55-'59), I used the Datatron 205 digital computer ( like the IBM 650). At Bell Telephone Labs ('61-'77) I and colleagues used the latest digital computers and image making equipment to study nonlinear waves and fluids.

9. One such wizard was Gary S. Deem. He came to Bell Labs as a pre-PhD and joined me in the research on nonlinear wave, Newtonian-fluid and chemical dynamics. We also simulated the Korteweg-de Vries ( KdV) nonlinear partial differential equation (pde) from which the *soliton* discovery emerged. He made the first animation of a solution of the KdV in 1965, now available at UTube.

10. Cosmic Vortex Projectiles. An artistic animation by Hilary Shames with Norman Zabusky, 2000.. https://www.youtube.com/watch?v=L7_pgPlpUIY. First shown at 3rd Science and Art Congress, ScArt3, Zurich 2000, as part of my , keynote talk "Scientific Computing Visualization - a new venue in the arts". See NZ.ScArt3.'00 , for a copy of the paper published in the Proceedings.

11. N.J. Zabusky organized , directed and contributed to ScArt4, the 4th Science and Art International Conference". ( New Brunswick, N.J. ,USA: 2005). (See http://mech2.rutgers.edu/scart4/a17links.htm. )

    N. J. ZAbusky was interviewed by Prof. Paul Leath, Chair of Department of Physics at Rutgers University. The mpg animation "From art to modern science: Understanding waves and turbulence" was shown on the Research Channel series , By the Book.'03 . The video was produced for the Research Channel by Rutgers,The State University of New Jersey, July 24, 2003.

12. In Figure 1, the two lowest panels in the column at right show instantaneous *images* from an evolving *3D vortex reconnection*; the second panel ( lower, L-to-R) shows *projections of integrated vorticity for this simulation at* four different instances. The lowest-left panel shows four instances of the evolution of 2D density (left) and vorticity (right) for the *shock-interface* (*Richtmyer-Meshkov)* environment .*Space-time projections* of "integrated-vorticity" are shown in the top three panels at left.

13. Visiometrics refers to: Visualization , quantification and modeling of evolving amorphous structures. This neologism was introduced in 1990 by F.J. Bitz and N.J. Zabusky in "DAVID and Visiometrics: Visualizing and quantifying evolving amorphous objects". Computers in Physics, Nov/Dec (603-614). Also see, R. Samtaney and N.J. Zabusky . High-Gradient compressible flows: Visualization, feature extraction and quantification" .In "Flow visualization, techniques and examples". Editors, A. Smits and T. Lim. Imperial College Press, 2000, 317-344. Also, visit the animation "VISIOMETRICS" , produced and narrated by Simon Cooper in mid-1990's at Rutgers

University, Laboratory for Visiometrics and Modeling. https://www.youtube.com/watch?v=B_jGsLMyHDo.

14. Braun, Marta. Picturing Time: The Work of Etienne-Jules Marey, 1830-1904.Chicago: University of Chicago Press, 1994. Marey was a French scientist, physiologist and chronophotographer. His last major works (1899-1902) were devoted to the observation and instant photography of narrow smoke layers moving around fixed objects, in one of the first modern aerodynamic wind tunnels. (Excerpted from Jules_Marey at Wikipedia) .

15. Marcel Duchamp (1887 – 1968) was a French artist who painted in a Cubist style, and added an impression of motion by using *repetitive imagery.* During this period Duchamp's fascination with transition, change, movement and distance became manifest, and like many artists of the time, he was intrigued with the concept of depicting a "Fourth dimension" in art. (Excerpted from WIKI).

16. Harold “Doc” Edgerton was MIT Institute Professor of Electrical Engineering. Photographs are from, “Stopping Time: The Photographs of Harold Edgerton”. Estelle Jussim (Author), Gus Kayafas (Editor), Harold Edgerton (Photographer) , 1987. See http://edgerton-digital-collections.org/galleries/iconic. Also see: Kris Belden-Adams , Harold Edgerton and complications of the 'Photographic Instant’. Frame Journal, See, http://framejournal.org/view-article/14 . Also see URL: http://www.agallery.com/pages/photographers/edgerton.html.

17. URL’s for some artists.

a) Marcel Christ. 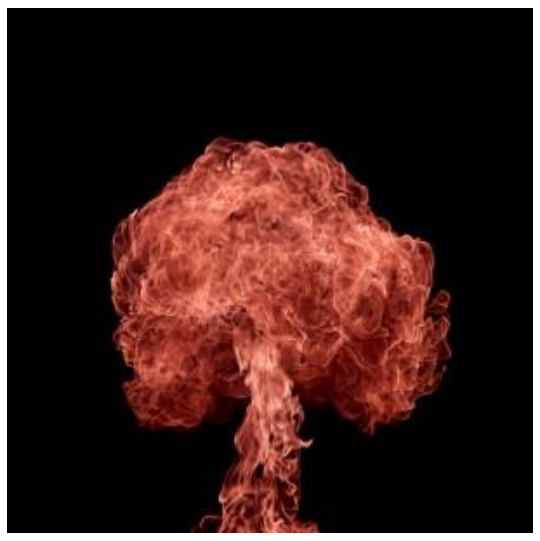 http://www.marcelchrist.com/non_commissioned;

b) Jack Long 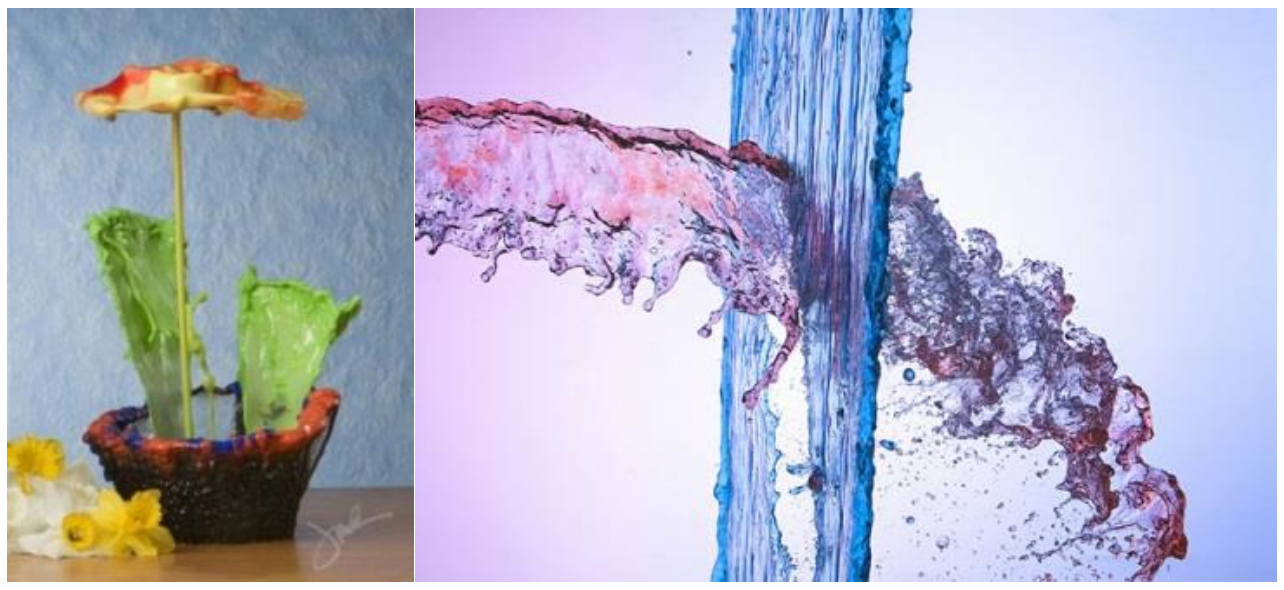

*LEFT :*http://www.slrlounge.com/jack-longs-water-curtain-tutorial.

*RIGHT:* http://www.cuded.com/2013/01/incredible-splash-photography-by-jack-long/

***END***